\documentclass[12pt]{article}

\usepackage{amssymb,amsfonts,amsthm}
\usepackage{amsmath}
\usepackage{bm}             
\usepackage[mathscr]{eucal}
\usepackage{cancel}
\usepackage{wasysym}
\usepackage{graphicx}
\usepackage{subfigure}
\def\be{\begin{equation}}
\def\ee{\end{equation}}
\def\bea{\begin{eqnarray}}
\def\eea{\end{eqnarray}}

\def\hsp5{\hspace{5mm}}

\theoremstyle{remark}

\newcommand{\sfrac}[2]{{\textstyle{#1\over#2}}}

\title{\sc Inflationary $\alpha$-attractor cosmology: A global dynamical systems perspective}

\begin{document}

\author{
\sc Artur Alho,$^{1}$\thanks{Electronic address:{\tt
aalho@math.ist.utl.pt}}\,  and Claes Uggla$^{2}$\thanks{Electronic
address:{\tt claes.uggla@kau.se}}\\
$^{1}${\small\em Center for Mathematical Analysis, Geometry and Dynamical Systems,}\\
{\small\em Instituto Superior T\'ecnico, Universidade de Lisboa,}\\
{\small\em Av. Rovisco Pais, 1049-001 Lisboa, Portugal.}\\
$^{2}${\small\em Department of Physics, Karlstad University,}\\
{\small\em Av. Universitetsgatan 1-2, S-65188 Karlstad, Sweden.}}

\date{}
\maketitle
%

\begin{abstract}

We study flat Friedmann-Lema\^itre-Robertson-Walker $\alpha$-attractor $\mathrm{E}$- and $\mathrm{T}$-models
by introducing a dynamical systems framework that yields regularized
unconstrained field equations on two-dimensional compact state spaces.
This results in both illustrative figures and a complete description of
the entire solution spaces of these models, including asymptotics. In
particular, it is shown that observational viability, which requires a
sufficient number of $e$-folds, is associated with a particular solution given
by a one-dimensional center manifold of a past asymptotic de Sitter state,
where the center manifold structure also explains why nearby solutions
are attracted to this ``inflationary attractor solution''. A center
manifold expansion yields a description of the inflationary regime with
arbitrary analytic accuracy, where the slow-roll approximation asymptotically
describes the tangency condition of the center manifold at the asymptotic
de Sitter state.

\end{abstract}

\section{Introduction}

Recently, there have been considerable developments as regards large field
inflation with plateau-like inflaton potentials, driven by the models
compatibility with observational data~\cite{plaXX15} and their ties to
supergravity and string theory~\cite{kallin3b}--\cite{lin16}. 
On the theoretical side it has, for example, been shown that such
models naturally arise from phenomenological supergravity. The underlying
hyperbolic geometry of the moduli space and the flatness of the K{\"a}hler
potential in the inflation direction is conveniently described in terms of
a field variable $\phi$, which gives rise to a kinetic term in the Lagrangian
with a pole at the boundary of the moduli space. By making a transformation
to a canonical variable $\varphi$, the moduli space near its boundary becomes
stretched, leading to an inflationary potential $V(\varphi)$ with an
asymptotic plateau-like form for $V(\varphi)$. This stretching results in predictions
that are quite insensitive to the original form of the potential $V(\phi)$.
In particular this leads to universal properties in $n_s-r$ diagrams, which
motivates calling these models $\alpha$-attractors, where $\alpha$ is a
constant parameter describing the exponential asymptotic flattening of the
potential $V(\varphi)$ in the Einstein frame. It should be noted that this
theoretical perspective unifies and contextualizes results for several previous
models such as the Starobinski and the Higgs inflation models. For additional
intriguing aspects such as couplings to other fields with couplings which
become exponentially small for large fields $\varphi$, as well as further
background discussions, see~\cite{kallin3b}--\cite{lin16}.

\begin{figure}[ht!]
\begin{center}
\subfigure[$V = V_0\left(1 - e^{-\sqrt{\frac{2}{3\alpha}}\varphi}\right)^{2n}$, with $n=\alpha=1$.]{\label{fig:potE}
\includegraphics[width=0.45\textwidth, trim = 0cm 0cm 0cm 0cm]{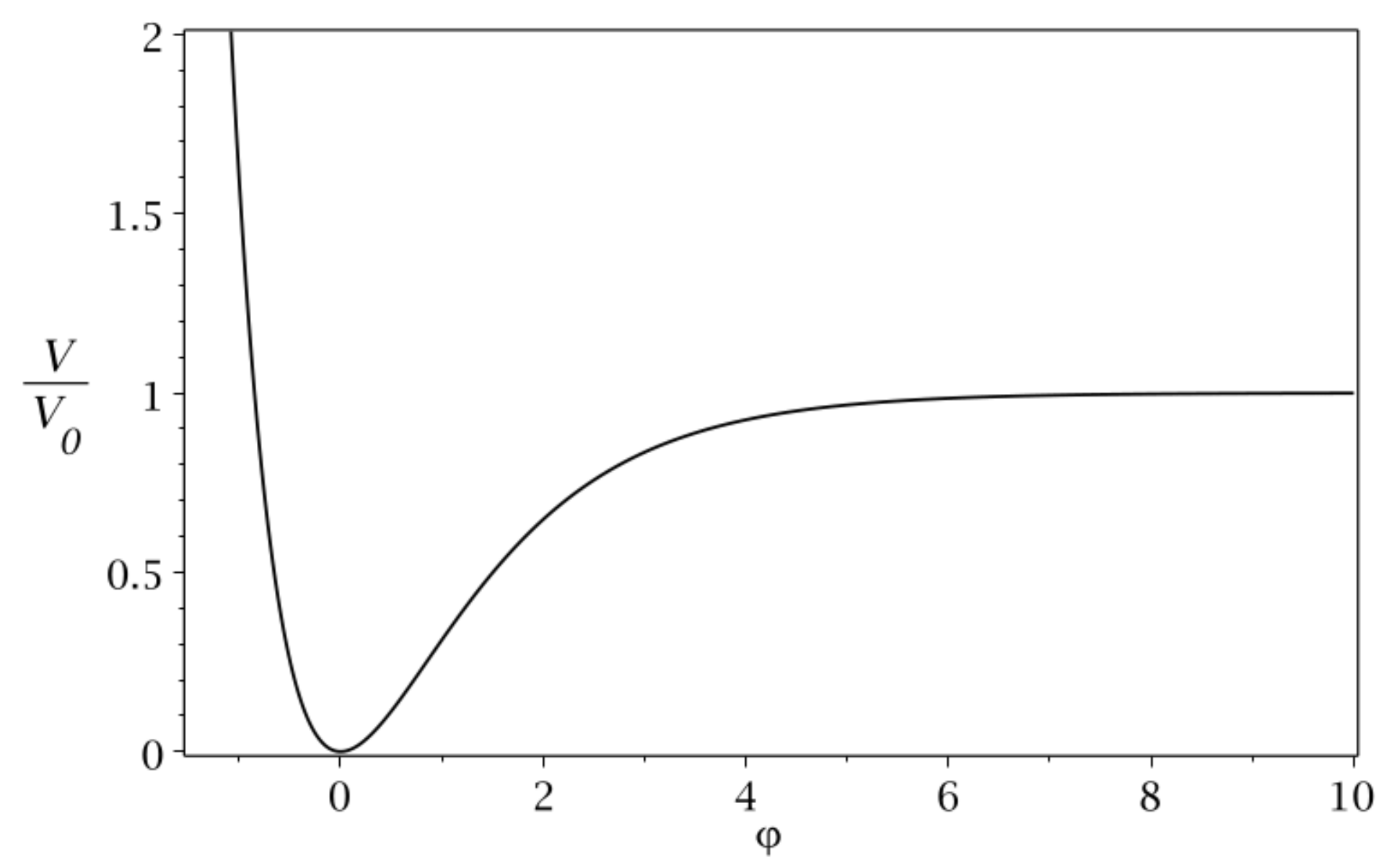}}
\subfigure[$V = V_0\tanh^{2n}\frac{\varphi}{\sqrt{6\alpha}}$, with $n=\alpha=1$]{\label{fig:potT}
\includegraphics[width=0.45\textwidth, trim = 0cm 0cm 0cm 0cm]{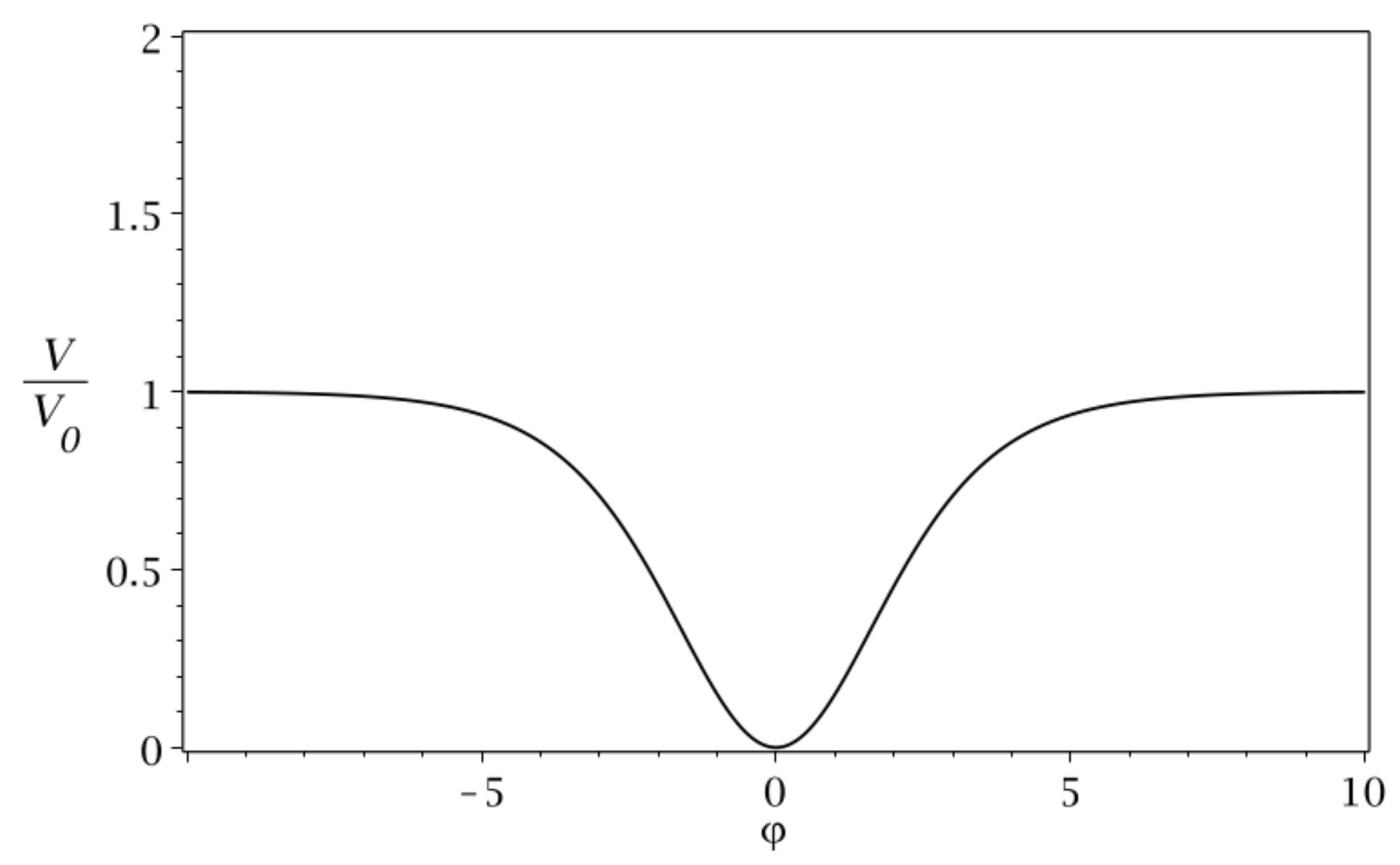}}
\end{center}
\vspace{-0.5cm}
\caption{The potentials of the $\mathrm{E}$- and $\mathrm{T}$-models.}
\label{fig:pot}
\end{figure}

We take the Einstein frame formulation as our starting point and restrict
the discussion to the flat, spatially homogeneous, isotropic
Friedmann-Lema\^{i}tre-Robertson-Walker (FLRW) spacetimes.
We thereby consider the following field equations for a canonically normalized
inflaton field $\varphi$ with a potential $V(\varphi)$:\footnote{We use reduced Planck
units: $c=1=8\pi G = 8\pi M_\mathrm{Pl}^{-2}$.}
\begin{subequations}\label{Fieldeq}
\begin{align}
3H^2 &= \sfrac12 \dot{\varphi}^2 + V(\varphi) = \rho_\varphi,\label{Gauss1}\\
\dot{H} &= -\sfrac{1}{2}\dot{\varphi}^2,\label{Ray1}\\
0 &= \ddot{\varphi} + 3H\dot{\varphi} + V_{\varphi},\label{KG}
\end{align}
\end{subequations}
where $V_{\varphi} = dV/d\varphi$; an overdot signifies the
derivative with respect to synchronous proper time, $t$; $H= \dot{a}/a$ is the Hubble
variable, where $a$ is the cosmological scale factor, with  
evolution equation $\dot{a} = aH$, which decouples from the above equations.
Furthermore, our focus will be on $\mathrm{E}$- and $\mathrm{T}$-models defined
by the potentials
\begin{subequations}\label{pot}
\begin{align}
V &= V_0\left(1 - e^{-\sqrt{\frac{2}{3\alpha}}\varphi}\right)^{2n},\label{Epot}\\
V &= V_0\tanh^{2n}\frac{\varphi}{\sqrt{6\alpha}},\label{Tpot}
\end{align}
\end{subequations}
with $V_0>0$, respectively (see Fig.~\ref{fig:pot}).

The modest aim in this paper is to present a dynamical systems formulation
that allows us to give a complete global classical description of the solution
spaces of the above models, including asymptotic properties, illustrated with
pictures describing compactified two-dimensional state spaces.
Our motivation for this is threefold:
\begin{itemize}
\item[(i)] As will be discussed, the above equations and potentials
allow one to rather directly get a feeling for the solution space and its
properties, but all aspects are not obvious. It is therefore of value to show
how a complete understanding can be obtained in a rigorous manner, especially
since this exemplifies how one can address other reminiscent problems.
\item[(ii)] By reformulating the problem as a regularized dynamical
system on a reduced compactified state space, we provide an example that
helps give the dynamical systems community access to inflationary cosmology.
This also makes powerful mathematical tools available to an area where such methods
have not, in our opinion, been used to full effect. For example, as will be shown,
the universal inflationary properties of the models are intimately connected with certain
aspects concerning center manifolds, which also result in approximation methods
complementing heuristic methods such as the slow-roll approximation.
\item[(iii)] The present dynamical systems formulation can be adapted so that 
it can be used in more general contexts such as anisotropic spatially 
homogeneous cosmology and even for models without any symmetries at all,
as shown for perfect fluids in e.g., Refs.~\cite{limetal04} and~\cite{ugg13a}.
Such an extension of the present paper would make it possible to address 
the issue of initial conditions for inflation in a generic context.
\end{itemize}

Let us now turn to the system~\eqref{Fieldeq}. By treating $\dot{\varphi}$ as
an independent variable, we obtain a reduced state space (because the equation for
the scale factor $a$ decouples) described by the state vector
$(H,\dot{\varphi},\varphi)$ obeying the constraint~\eqref{Gauss1}; i.e.,
the problem can be regarded as a two-dimensional dynamical system.
The $\mathrm{E}$- and $\mathrm{T}$-models are examples of models with a non-negative
potential $V(\varphi)$ with a single extremum point, a minimum, conveniently located at
$\varphi=0$, for which $V(0)=0$. As a consequence these models admit a Minkowski
(fixed point/critical point/equilibrium point)
solution at $(H,\dot{\varphi},\varphi) = (0,0,0)$. Due to~\eqref{Gauss1}, all other solutions
either have a positive or negative $H$. Since we are interested in cosmology we consider $H>0$.
It follows that $H$ is monotonically decreasing since Eq.~\eqref{Ray1} yields $\dot{H} \leq 0$,
where $\dot{H} =0$ requires $\dot{\varphi}=0$; moreover, $\dot{\varphi}$ cannot remain zero since
$\dot{\varphi}=0$ implies that $\ddot{\varphi}|_{\dot{\varphi}=0} = - V_\varphi$, where
$V_\varphi \neq 0$ since we exclude the only extremum point at
$(H,\dot{\varphi},\varphi) =(0,0,0)$  by assuming that $H>0$. Thus, the graph of $H$ just passes
through an inflection point when $\dot{\varphi}=0$.

The system~\eqref{Fieldeq} can be discussed in terms of the following heuristic picture: 
Equation~\eqref{KG} can be interpreted as an equation for a particle of unit mass 
with a one-dimensional coordinate $\varphi$, moving in a
potential $V(\varphi)$ with a friction force $-3H\dot{\varphi}$, while $3H^2$ can be viewed as
a monotonically decreasing energy in the constraint~\eqref{Gauss1}. Running backwards in time 
leads to a particle with ever-increasing energy moving in the potential $V(\varphi)$,
where $3H^2 \rightarrow \infty$. In combination with the shape of a given potential, this yields an
intuitive picture of the dynamics. Let us now turn to the specific potentials~\eqref{pot}. In the
case of the $\mathrm{E}$- and $\mathrm{T}$-models both potentials behave as
$\sim \varphi^{2n}$ for small $\varphi$. The potential for the $\mathrm{E}$-models
($\mathrm{T}$-models) has a plateau given by $V_0$ when $\varphi \rightarrow +\infty$
($\varphi \rightarrow \pm\infty$), while $V$ behaves as
$V_0\exp(-2n\sqrt{2/3\alpha}\,\varphi)$ when $\varphi \rightarrow -\infty$.
In contrast to the $\mathrm{E}$-models, the potential, and thereby the field equations,
of the $\mathrm{T}$-models exhibits a discrete symmetry under the transformation
$\varphi \rightarrow - \varphi$.

Let us begin our heuristic discussion of the dynamics of the $\mathrm{E}$- and
$\mathrm{T}$-models with dynamics toward the future. In both cases, if a solution
has obtained an ``energy'' $3H^2<V_0$ it follows straightforwardly that the solution
will end up at the Minkowski fixed point. But does it do so by just ``gliding'' down
the potential, or does it do so by damped oscillations, i.e., is the motion
of $\varphi$ overdamped or underdamped? This does not follow from the heuristic particle
discussion, but requires further analysis. Attempting to do so by solving the 
constraint~\eqref{Gauss1} for $H$, i.e., by setting 
$H= \sqrt{\dot{\varphi}^2/2 + V(\varphi)}/\sqrt{3}$ in~\eqref{KG}, leads to an
unconstrained two-dimensional dynamical system for $(\dot{\varphi},\varphi)$.
This system, however, has an unbounded state space and differentiability problems at 
$\varphi = 0$ for potentials that behave as $\sim\varphi^{2n}$ for small $\varphi$. 
These problems are of course not insurmountable, but they prevent 
global pictures of the solution spaces that accurately
reflect the asymptotic features of the solutions. Nevertheless, it is not difficult 
to show that the motion is underdamped and that solutions with $3H^2<V_0$ 
undergo damped oscillations. But do all solutions end at the Minkowski state, 
or are there solutions that asymptotically end with an energy $V_0$ at infinitely 
large $\varphi$? In other words, is the Minkowski state a global future attractor? (As we 
will see, the answer is ``yes'' for the present potentials.) One would also perhaps guess that there 
is a single solution that begins at an infinite value of the scalar field with an 
initial energy $3H^2 = V_0$, corresponding to an asymptotic de Sitter state, 
with the solution initially slowly rolling down the potential, which, as we will show, 
is indeed the case.

What about the past dynamics? From an inflationary point of view one would perhaps argue that
solutions are only physically acceptable with initial data for which $3H^2 \approx V_0$.
However, all solutions, except for the single one coming from the
de Sitter plateau with an initially infinite scalar field, originate from
$3H^2 \rightarrow \infty$. One might then take the viewpoint that the previous evolution
of solutions before $3H^2 \approx V_0$ is physically irrelevant. However, all this assumes
that the inflationary scenario is correct and prevents investigations that either justify 
this or cast doubt on it. Moreover, even if this is assumed to be correct, one still wants
approximations for the solutions before the quasi-de Sitter stage, as done
in, e.g., Ref.~\cite{caretal15a}, where a massless state is used for this purpose. But this is
intimately connected with the limit $3H^2 \rightarrow \infty$, so let us therefore consider
this limit from the present heuristic perspective.

Because the potential~\eqref{Tpot} is symmetric and bounded, the heuristic description of the dynamics
of the $\mathrm{T}$-models is simpler than that for the $\mathrm{E}$-models. We therefore begin 
with the $\mathrm{T}$-models. Since the potential in this case is bounded, $V(\varphi) \leq V_0$,
it follows from~\eqref{Gauss1} that the limit $3H^2 \rightarrow \infty$ implies that 
$\dot{\varphi}^2/2V(\varphi) \rightarrow 0$ and hence that the influence of the potential on 
the dynamics becomes asymptotically negligible; i.e., the asymptotic state is expected to be
that of a massless scalar field. Furthermore, we expect two physically equivalent representations
associated with $\varphi \rightarrow \pm \infty$ toward the past. 

For the $\mathrm{E}$-models the potential is no longer bounded
and this results in a somewhat more complicated situation. Nevertheless, toward the past, 
irrespective of the value of $\alpha$, we would expect an open set of models to approach a
massless state toward $\varphi \rightarrow +\infty$ for similar reasons as for the 
$\mathrm{E}$-models. Furthermore, we expect that whether or not all solutions end up there depends
on the steepness of the potential in the limit $\varphi \rightarrow -\infty$. In this limit
the potential behaves as an exponential, and problems with an exponential can be viewed as
scattering a particle against a potential wall if the potential is steep enough; in
such a situation we expect that the massless state with $\varphi \rightarrow +\infty$ 
describes the past asymptotic behavior of all solutions. 
If the potential is not steep enough to accomplish this the situation becomes 
more complicated. Nevertheless, based on the knowledge of the dynamics of a single
exponential one might guess that there is an open set of solutions that approaches a
massless state at $\varphi \rightarrow - \infty$ and that there exists a single solution 
that approaches this limit in a power-law fashion. Moreover, if the steepness of the 
exponential limit is sufficiently moderate, we expect this solution to describe an
early inflationary power-law state.

Although quite helpful, the above heuristic considerations lead to a rather scattered 
impression of the global dynamics and leave some remaining unanswered questions. In addition,
we have not obtained any quantitative results, and, what is even worse, the heuristic
picture runs into increasing problems when trying to use it in more general contexts involving
additional degrees of freedom. In contrast, the dynamical systems formalism presented below is
applicable to more general situations than the present one, which rather acts as a
pedagogical example. With this in mind we will now develop a formalism that
addresses the above issues and at a glance describes the global situation rigorously, and also
makes techniques available for describing quantitatively correct approximations for solutions
in the past and future regimes (indeed, they can be combined to even give global approximations
for the solutions to arbitrary accuracy if one is so inclined, see~\cite{alhugg15}). 

To avoid differentiability problems and to obtain asymptotic approximations and a global 
picture of the solution space, we change variables. We do so by
following the treatment of monomial potentials $V \propto \varphi^{2n}$
in~\cite{alhugg15} and~\cite{alhetal15}, but adapting the formulation to the
particular features of the potentials~\eqref{Epot} and~\eqref{Tpot}.\footnote{To treat models
with positive potentials, see e.g.~\cite{ugg13} and~\cite{alhugg15b}; for examples
of other work on scalar fields using dynamical systems methods, see e.g.~\cite{hal87}--\cite{fadetal14}.
For a recent rather general discussion on dynamical systems formulations and methods in other
cosmological contexts, see~\cite{alhetal16}.} We derive three complementary dynamical
systems since the global one is not optimal for quantitative descriptions in all parts
of the state space; instead the global picture should be seen as a collecting ground which
gives the overall picture. Since the dependent variables are defined in a similar manner for 
$\mathrm{E}$- and $\mathrm{T}$-models, while the independent variables differ, we define the 
three complementary sets of dependent variables in this section, while the independent variables and
the dynamical systems and their analysis will be presented in the subsequent two sections.

We begin by defining the Hubble-normalized or, equivalently in the present flat FLRW cases,
energy density-normalized dimensionless variables (thereby capturing the physical essence
of the problem), by making the following variable transformation,
$(H,\dot{\varphi},\varphi) \rightarrow (\tilde{T},\Sigma_\varphi,X)$~\footnote{The notation
$\Sigma_\varphi$ is used because $\Sigma_\varphi$, in a multidimensional Kaluza-Klein
perspective, is analogous to Hubble-normalized shear in anisotropic cosmology,
where $\Sigma$ is standard notation; see e.g. Ref.~\cite{waiell97}.}:
\begin{subequations}\label{vardef}
\begin{align}
\tilde{T} &= \left(\frac{V_0}{3H^2}\right)^{\frac{1}{2n}} =
\left(\frac{V_0}{\rho_\varphi}\right)^{\frac{1}{2n}}, \label{tildeTdef}\\
\Sigma_\varphi &= \frac{1}{\sqrt{6}}\frac{\dot{\varphi}}{H} =
\frac{1}{\sqrt{6}}\frac{d\varphi}{dN}, \label{Sigdef}\\
X &= \left(\frac{V(\varphi)}{3H^2}\right)^{\frac{1}{2n}} =
\left(\frac{V(\varphi)}{\rho_\varphi}\right)^{\frac{1}{2n}}, \label{Xdef}
\end{align}
where $X$ takes the following explicit form for the $\mathrm{E}$- and $\mathrm{T}$-models,
\begin{align}
X &= \left(\frac{V_0}{3H^2}\right)^{\frac{1}{2n}}\left(1 - e^{-\sqrt{\frac{2}{3\alpha}}\varphi}\right)
= \tilde{T}\left(1 - e^{-\sqrt{\frac{2}{3\alpha}}\varphi}\right), \label{XdefE}\\
X &= \left(\frac{V_0}{3H^2}\right)^{\frac{1}{2n}}\tanh\frac{\varphi}{\sqrt{6\alpha}}
= \tilde{T}\tanh\frac{\varphi}{\sqrt{6\alpha}}, \label{XdefT}
\end{align}
\end{subequations}
respectively. The quantity $N = \ln(a/a_0)$ in~\eqref{Sigdef} represents the number of $e$-folds
with respect to some reference time $t_0$ where $a(t_0) = a_0$. Below, for simplicity, we 
assume that $n$ is a positive integer.

Since $H$ is monotonically decreasing it follows that $\tilde{T}$ is monotonically increasing.
Furthermore, when $H\rightarrow 0\,  \Rightarrow\, \tilde{T} \rightarrow \infty$,
$H\rightarrow \infty\, \Rightarrow\, \tilde{T} \rightarrow 0$, and 
$3H^2 \rightarrow V_0\, \Rightarrow\, \tilde{T} \rightarrow 1$. With the above definitions, the
Gauss constraint~\eqref{Gauss1} takes the form
\begin{equation}
1 = \Sigma_\varphi^2 + X^{2n}.
\end{equation}
The state space thereby has a cylinderlike structure (cylinder structure when $n=1$).
The constraint can be solved globally by introducing an angular variable $\theta$
according to
\begin{subequations}\label{thetadef}
\begin{align}
\Sigma_\varphi &= G(\theta) \sin\theta , \qquad X = \cos\theta, \\
G(\theta) &= \sqrt{\frac{1-\cos^{2n}\theta}{1-\cos^2\theta}} = \sqrt{\sum_{k=0}^{n-1}\cos^{2k}\theta} ,\label{Gdef}
\end{align}
\end{subequations}
which leads to an unconstrained dynamical system for the state vector $(\tilde{T},\theta)$.
For future purposes we note that $G \geq 1$ (with $G\equiv 1$ when $n=1$), and
\begin{equation}
G(0) = \sqrt{n}.
\end{equation}
To obtain a bounded (relatively compact) state space
with state vector $(T,\theta)$, we make the transformation
\begin{equation}\label{Tdef}
T = \frac{\tilde{T}}{1 + \tilde{T}}, \qquad \tilde{T} = \frac{T}{1-T}.
\end{equation}
Thus, $T$ is monotonically increasing where $H\rightarrow 0\, \Rightarrow\, T \rightarrow 1$, 
$H\rightarrow \infty\, \Rightarrow\, T \rightarrow 0$, and 
$3H^2 \rightarrow V_0\, \Rightarrow\, T \rightarrow \frac12$.

The deceleration 
parameter, $q$, 
is defined and given by
\begin{equation}
q = -\frac{\ddot{a}}{aH^2} = - (1 + H^{-2}\dot{H}) = - 1 + 3\Sigma_\varphi^2 = 2 - 3\cos^{2n}\theta ,
\end{equation}

To proceed with the choice of independent variables, we treat the
$\mathrm{E}$- and $\mathrm{T}$-models separately in the following two
sections, where we also perform a complete local and global
dynamical systems analysis of these models. We end the paper with
some concluding remarks in Sec.~\ref{sec:concl}, e.g. about the
relationship between the center manifold analysis performed in the
two $\mathrm{E}$- and $\mathrm{T}$-model sections and the slow-roll
approximation.

\section{$\mathrm{E}$-models}

\subsection{Dynamical systems formulations}

Using the dependent variables given in Eq.~\eqref{vardef} for the
$\mathrm{E}$-models and $N=\ln(a/a_0)$ as the independent variable, where
\begin{equation}
\frac{dN}{dt} = H,
\end{equation}
results in the following evolution equations for the state vector
$(\tilde{T},\Sigma_\varphi,X)$,
\begin{subequations}\label{dynsys1E}
\begin{align}
\frac{d\tilde{T}}{dN} &= \frac{3}{n}\Sigma_\varphi^2\tilde{T},\label{tildeTSE}\\
\frac{d\Sigma_\varphi}{dN} &= -3\left(\Sigma_\varphi X + \bar{\lambda}(\tilde{T} - X)\right)X^{2n-1},\\
\frac{dX}{dN} &= \frac{3}{n}\left(\Sigma_\varphi X + \bar{\lambda}(\tilde{T} - X)\right)\Sigma_\varphi,
\end{align}
and the constraint
\begin{equation}\label{Gauss2}
1 = \Sigma_\varphi^2 + X^{2n},
\end{equation}
\end{subequations}
where it has been convenient to define
\begin{equation}
\bar{\lambda} = \frac{2n}{3\sqrt{\alpha}}.
\end{equation}

The state space is bounded by the conditions that $\tilde{T}>0$ and that
\begin{equation}\label{phi-tildeT}
\tilde{T} e^{-\sqrt{\frac{2}{3\alpha}}\varphi} = \tilde{T} - X >0.
\end{equation}
Since
\begin{equation}
\frac{d}{dN}(\tilde{T} - X ) = \frac{3}{n}\left(\Sigma_\varphi - \bar{\lambda}\right)\Sigma_\varphi(\tilde{T} - X),
\end{equation}
it follows that the physical state space is bounded toward the past by the invariant subsets
$\tilde{T}=0$ for $X \leq 0$ and $\tilde{T} - X =0$ for $X \geq 0$ (recall that
$\tilde{T}$ is monotonically increasing toward the future and therefore decreasing
toward the past). Their intersection, $\Sigma_\varphi = \pm 1, X=0$, corresponds to
two fixed points $\mathrm{M}_\pm$. Furthermore, the $\tilde{T} - X =0$ subset is
divided into two disconnected parts by a fixed point $\mathrm{dS}$ located at $\tilde{T}=1=X$.
Due to the regularity of the above equations we can include the invariant boundary,
($\tilde{T}=0$ for $X \leq 0$) $\cup$ ($\tilde{T} - X =0$ for $X \geq 0$), which
we refer to as the past boundary. Indeed, it is necessary to include
this boundary in order to describe past asymptotics since, as will be shown, all
solutions originate from the fixed points on this boundary. Note further that the
equations on the $\tilde{T}=0$ subset are identical to those for an exponential
potential $V = V_0e^{-\sqrt{6}\bar{\lambda}\varphi}$ since in this case
%
\begin{equation}
\frac{d\Sigma_\varphi}{dN} = -3(\Sigma_\varphi - \bar{\lambda})X^{2n}
= -3(\Sigma_\varphi - \bar{\lambda})(1- \Sigma_\varphi^2). 
\end{equation}
%
Moreover, the equations on the $\tilde{T}=X$ subset are identical to those that
correspond to a constant potential, which can be seen by setting $\bar{\lambda}=0$
in the above equation.

By globally solving the constraint by using Eq.~\eqref{thetadef} we
obtain the following unconstrained dynamical system
\begin{subequations}\label{dynsys2E}
\begin{align}
\frac{d\tilde{T}}{dN} &= \frac{3}{n}\left(1 - \cos^{2n}\theta\right)\tilde{T},\label{tildeTthE}\\
\frac{d\theta}{dN} &= -\frac{3}{2n}\left(G\sin2\theta + 2\bar{\lambda}(\tilde{T} - \cos\theta)\right)G.
\end{align}
\end{subequations}
%

\begin{figure}[ht!]
\begin{center}
\subfigure[$0< \bar{\lambda} <1$]{\label{fig:SSE_wPL}
\includegraphics[width=0.45\textwidth]{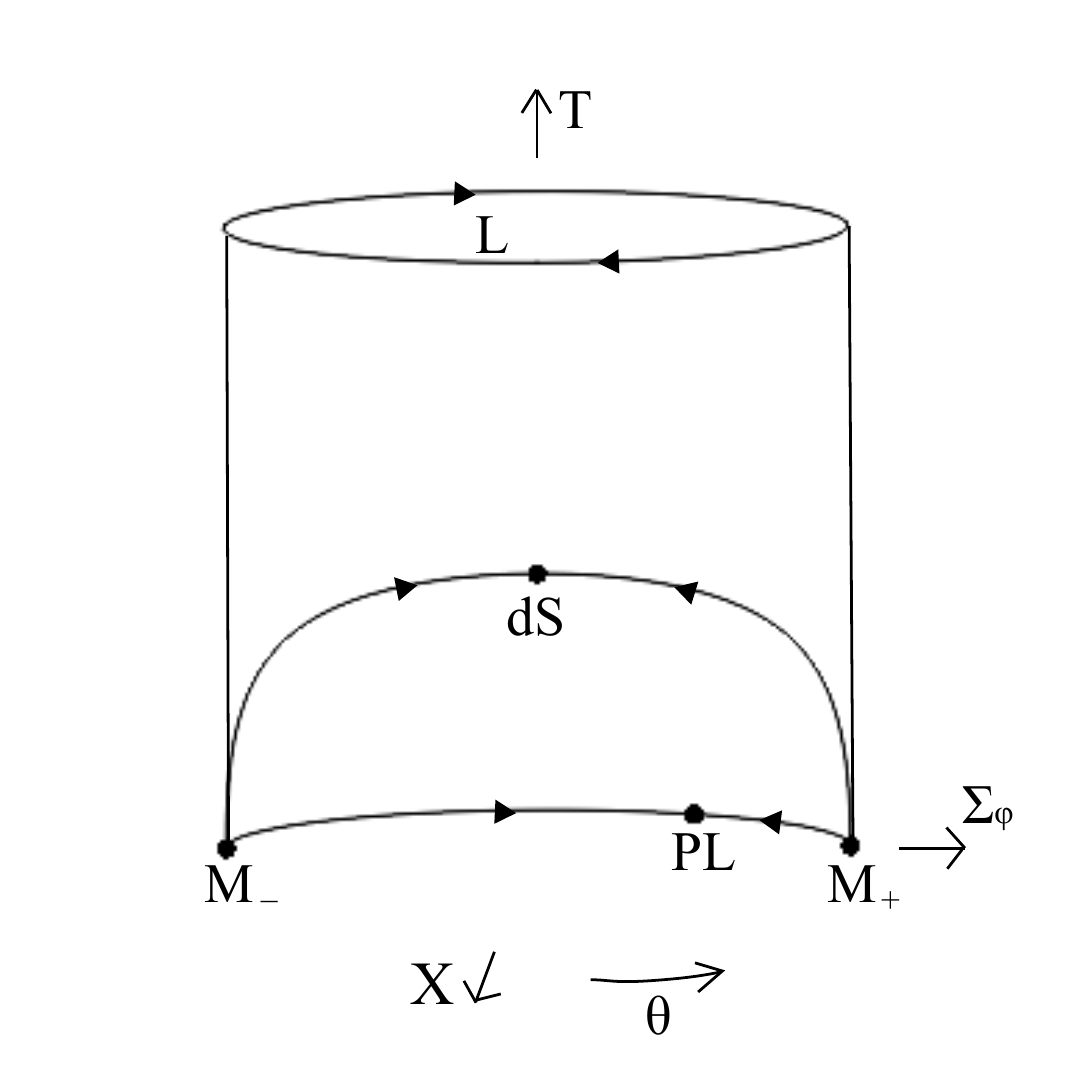}}
\subfigure[$\bar{\lambda} \geq 1$]{\label{fig:SSE_noPL}
\includegraphics[width=0.45\textwidth]{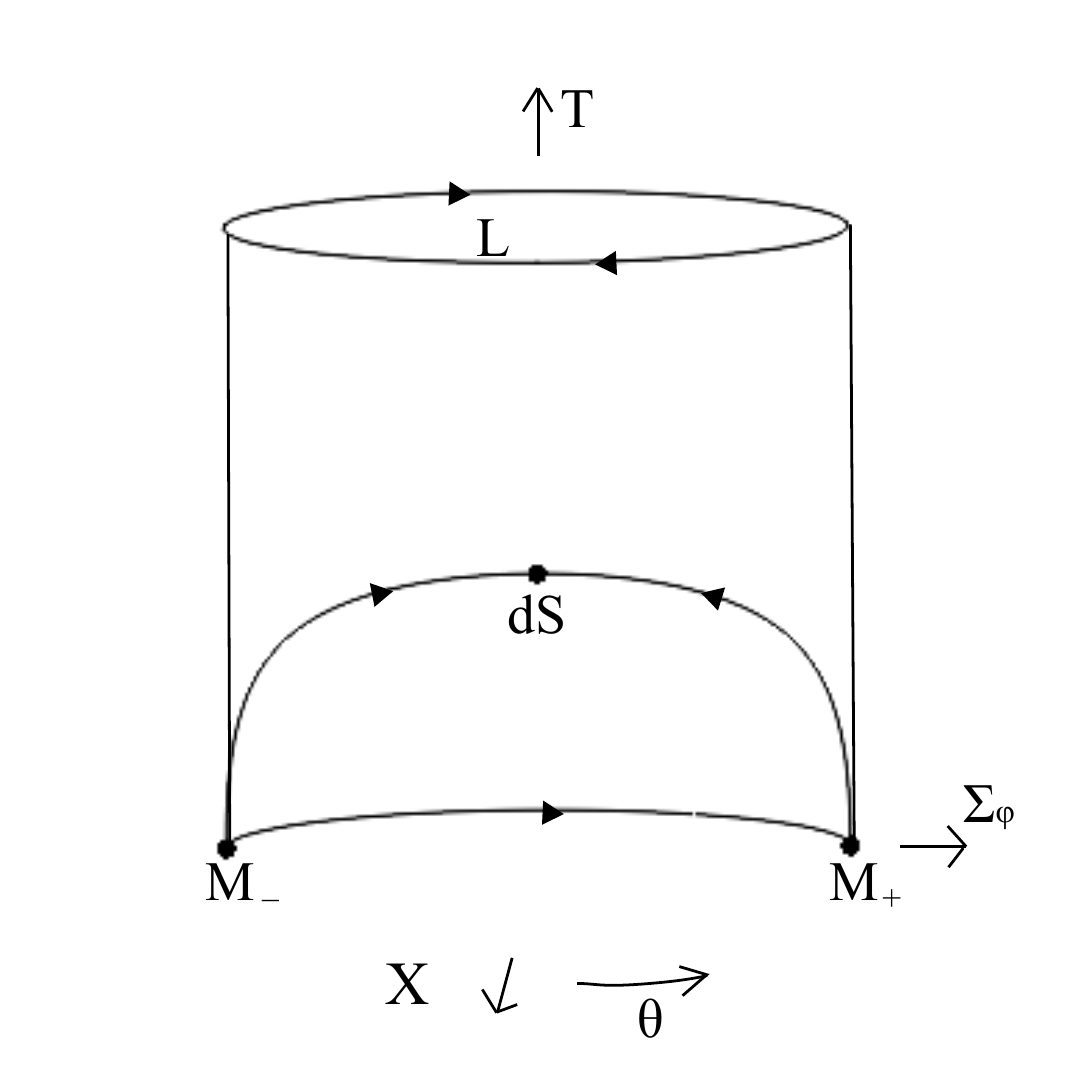}}
\end{center}
\vspace{-0.5cm}
\caption{State space and boundary structures for the $\mathrm{E}$-models with
$V = V_0\left(1 - e^{-\sqrt{\frac{2}{3\alpha}}\varphi}\right)^{2n}$.
Recall that $\bar{\lambda} = \frac{2n}{3\sqrt{\alpha}}$. 
}\label{fig:Estatespace}
\end{figure}

Finally, by using $T$ and $\theta$ and changing the time variable from $N$ to $\bar{\tau}$
according to
\begin{equation}
\frac{d\bar{\tau}}{dN} = 1+\tilde{T} = \frac{1}{1-T},
\end{equation}
we obtain the regular dynamical system
\begin{subequations}\label{dynsys3E}
\begin{align}
\frac{dT}{d\bar{\tau}} &= \frac{3}{n}T(1-T)^2(1 - \cos^{2n}\theta),\label{TthE}\\
\frac{d\theta}{d\bar{\tau}} &= -\frac{3}{2n}\left[(1-T)G \sin2\theta  + 2\bar{\lambda}\left(T-(1-T)\cos{\theta}\right)\right]G.
\end{align}
\end{subequations}
Apart from including the past boundary, which in the present variables is given by
$T=0$ when $\cos\theta \leq 0$ and $T - (1-T)\cos\theta =0$ for $\cos\theta\geq 0$, we also include the 
future boundary $T=1$, which corresponds to $H=0$ and the final Minkowski state. Thus,
the resulting extended state space is given by a finite cylinder with the region
$T - (1-T)\cos\theta < 0$ with $\cos\theta >0$ removed (see Fig.~\ref{fig:Estatespace}).
%

\subsection{Dynamical systems analysis}

\begin{figure}[ht!]
\begin{center}
\subfigure[Solution space for $0<\bar{\lambda}<1$.]{\label{fig:EPL}
\includegraphics[width=0.45\textwidth]{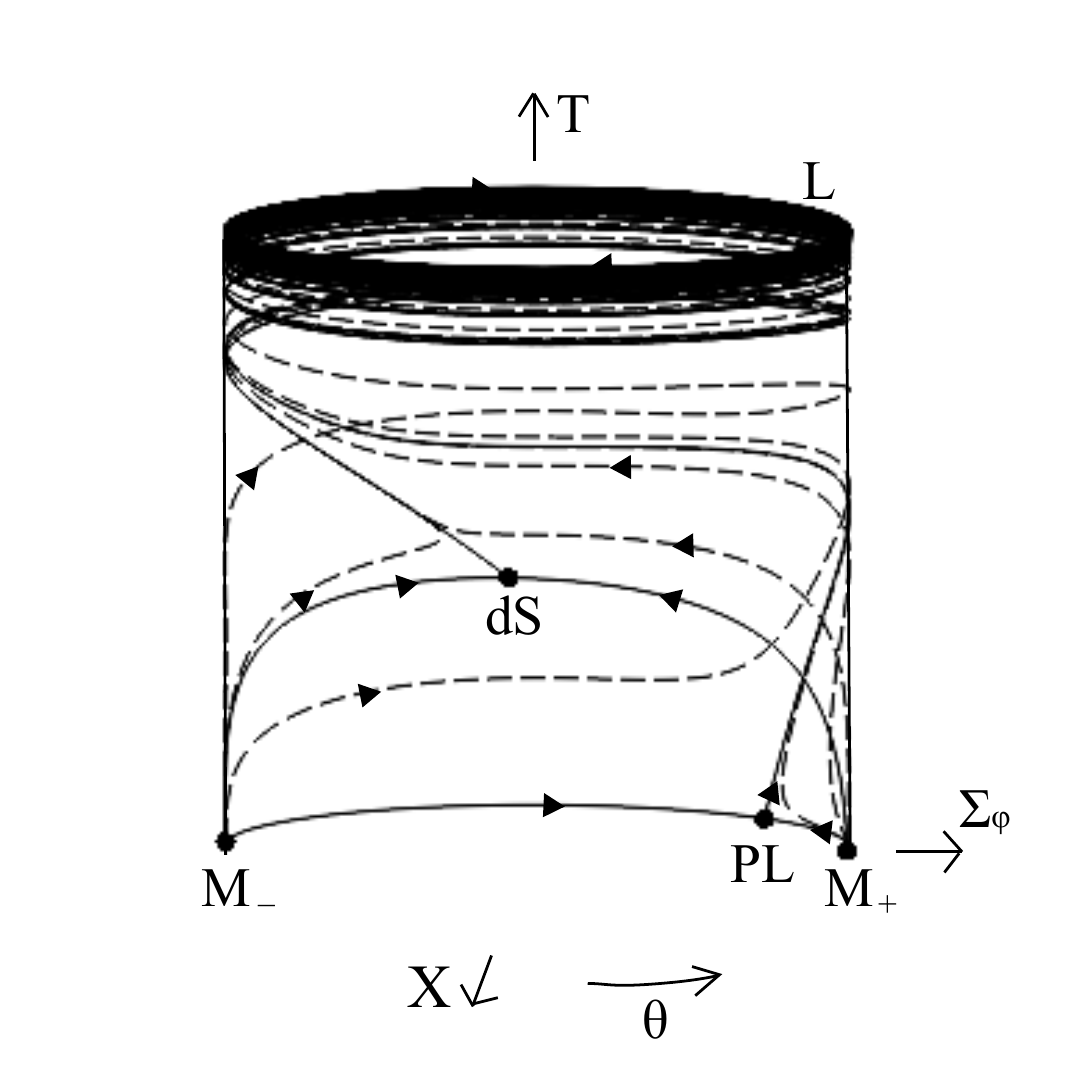}}
\subfigure[`Unwrapped' solution space for $0<\bar{\lambda}<1$.]{\label{fig:EPLunrapped}
\includegraphics[width=0.45\textwidth]{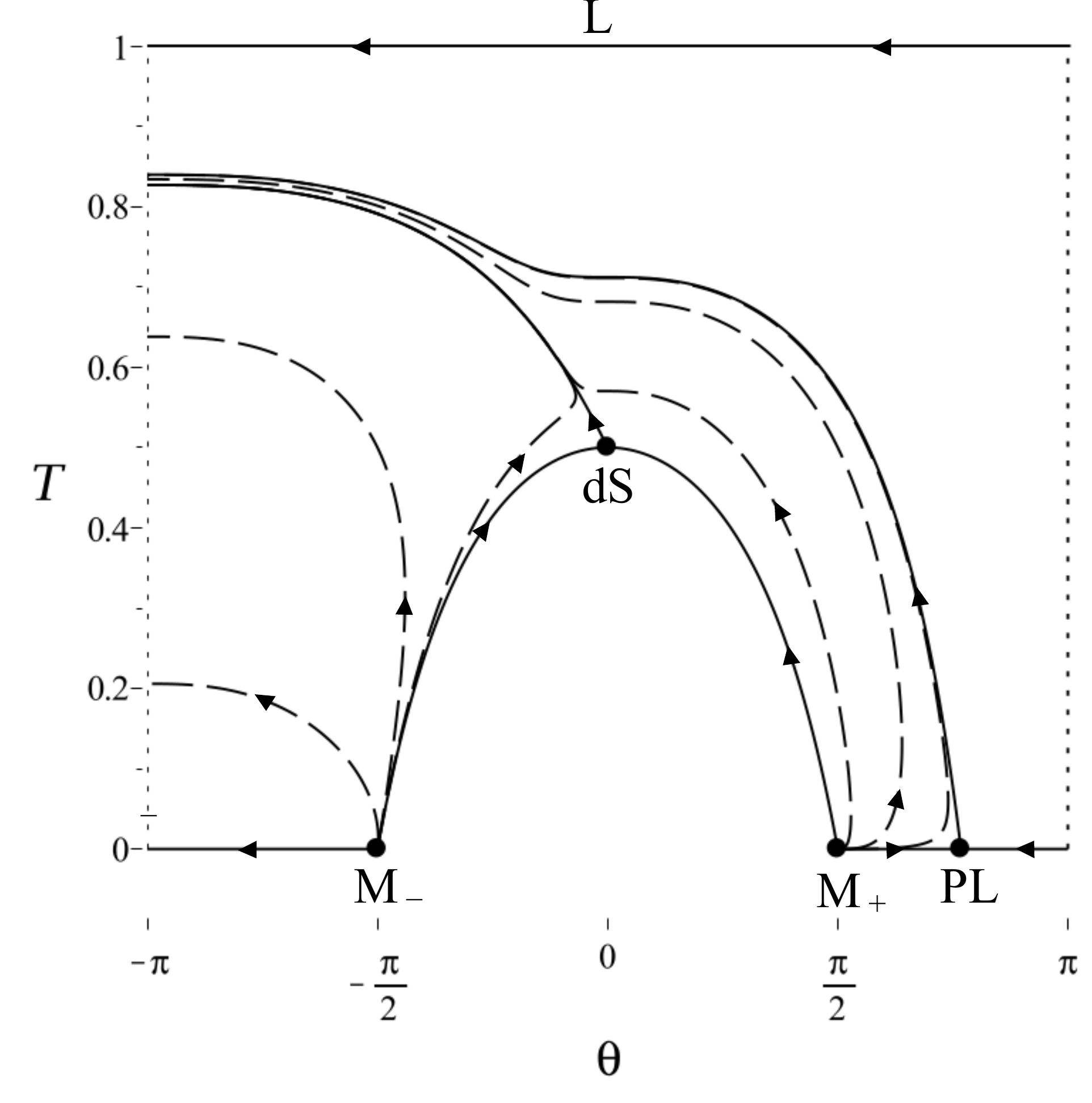}}
\subfigure[Solution space for $\bar{\lambda}\geq 1$.]{\label{fig:nonEPL}
\includegraphics[width=0.45\textwidth]{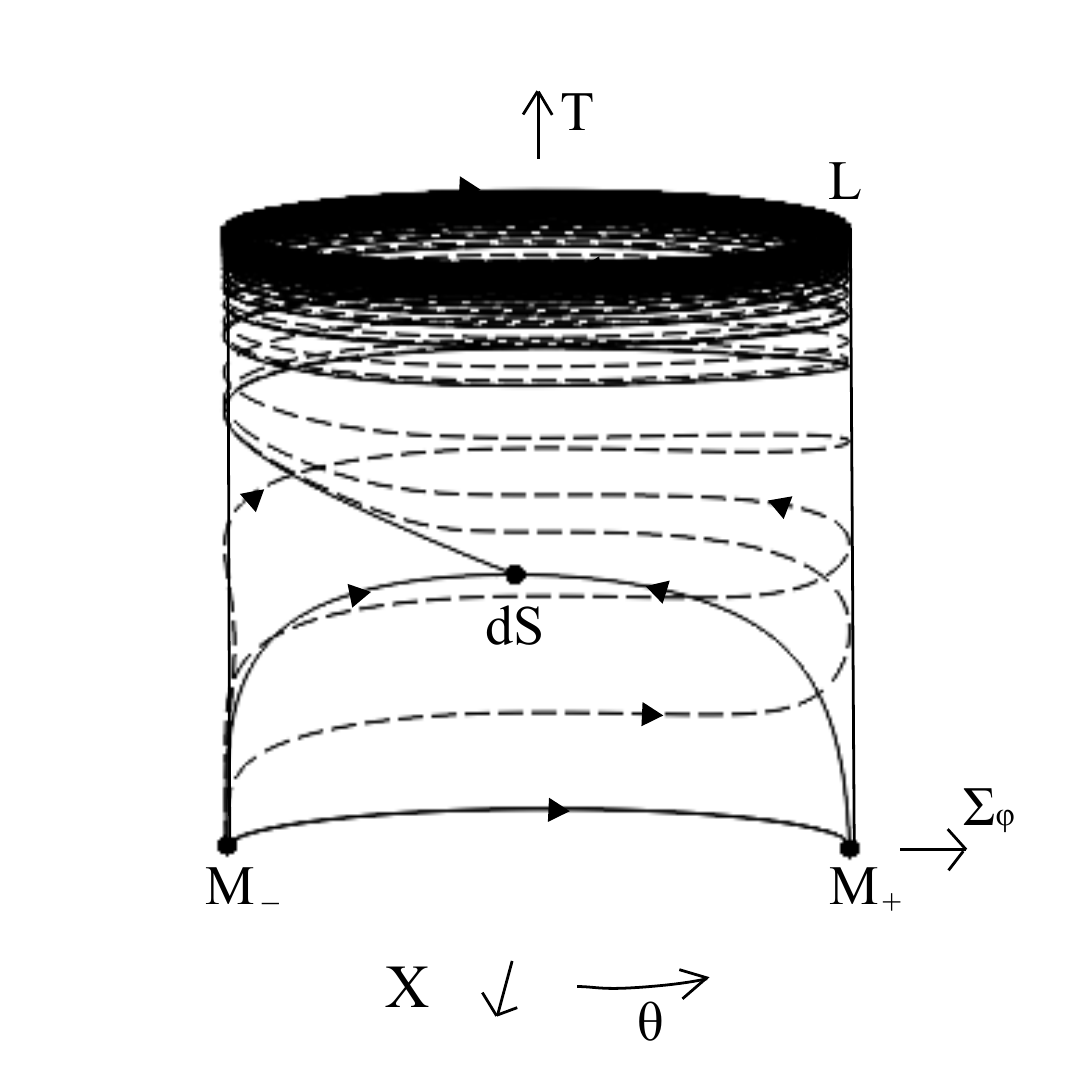}}
\subfigure[`Unwrapped' solution space for $\bar{\lambda}\geq 1$.]{\label{fig:nonEPLunwrapped}
\includegraphics[width=0.45\textwidth]{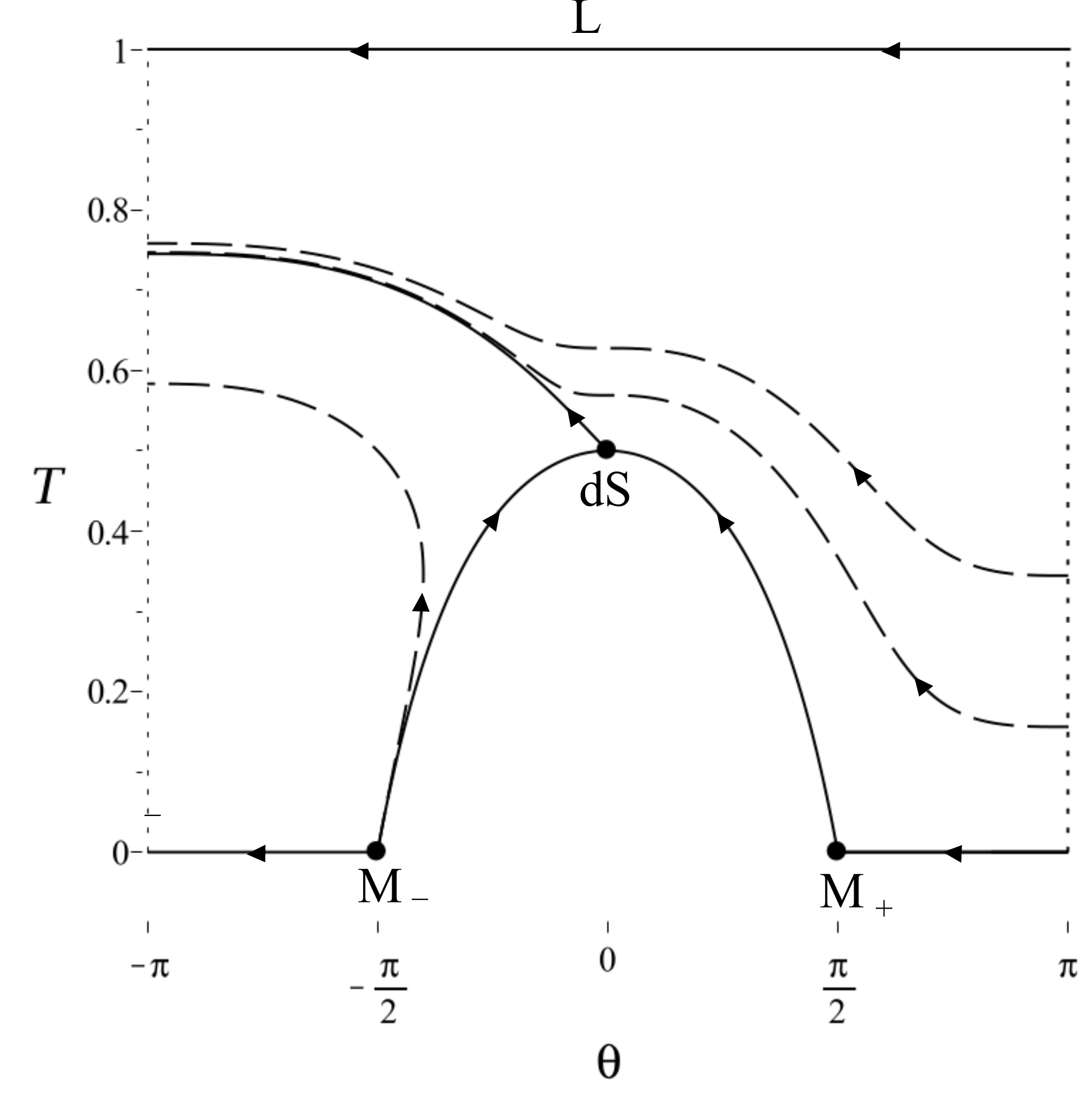}}
\end{center}
\vspace{-0.5cm}
\caption{Representative solutions describing the solution spaces for
$\mathrm{E}$-models with
$V = V_0\left(1 - e^{-\sqrt{\frac{2}{3\alpha}}\varphi}\right)^{2n}$.
In panels (a) and (b) $0<\bar{\lambda} <1$, which corresponds to $\alpha > (2n/3)^2$, represented
by the values $n=1$ and $\alpha=1$. In panels (c) and (d), $\bar{\lambda}\geq 1$, which
corresponds to $\alpha \leq (2n/3)^2$, represented by $n=1$ and $\alpha=1/4$.
}
\label{fig:Esolspace}
\end{figure}

From the definitions, and the fact that $H$ is monotonically decreasing,
it follows that $\tilde{T}$ and $T$ are monotonically increasing. This is also seen
in Eqs.~\eqref{tildeTSE}, \eqref{tildeTthE}, and \eqref{TthE}, although further insight
is gained by considering how $\tilde{T}$, and hence $T$, behaves
when $\Sigma_\varphi=0\, \Rightarrow\, \theta = m\pi$, where $q=-1$:
\begin{equation}\label{mon}
\left. \frac{d\tilde{T}}{dN}\right|_{q=-1} = 0,\qquad
\left. \frac{d^2\tilde{T}}{dN^2}\right|_{q=-1} = 0,\qquad
\left. \frac{d^3\tilde{T}}{dN^3}\right|_{q=-1} = \frac{54\bar{\lambda}^2}{n}(\tilde{T} - \cos(m\pi))^2\tilde{T}.
\end{equation}
Since $\tilde{T} > 1$ when $m$ is even and $\tilde{T}> 0$ when $m$ is odd in the
physical state space, it follows that by viewing the above as the coefficients
in a Taylor expansion, $\tilde{T}$, and hence $T$, is monotonically increasing, although
the graphs of $\tilde{T}$ and $T$ go through inflection points when $q=-1$. Furthermore, since
\begin{equation}\label{mon}
\left. \frac{d\theta}{dN}\right|_{q=-1} = -\frac{3\bar{\lambda}}{\sqrt{n}}\left(\tilde{T} - \cos(m\pi)\right),
\end{equation}
it follows that $\theta$ is monotonically decreasing at $q=-1$ and thus that the solution
curves in the $T,\theta$ state space become horizontal in $T$ at $q=-1$ (see
Fig.~\ref{fig:Esolspace}).

The monotonicity properties of $T$ show that there are no fixed points or recurring orbits
in the physical state space (i.e., the extended state space with the future and past invariant boundaries excluded).
All orbits originate from the past boundary and end at the future boundary at $T=1$, where
\begin{equation}\label{periodicLE}
\left. \frac{d\theta}{d\bar{\tau}}\right|_{T=1} = -\frac{3\bar{\lambda}}{n}G <0;
\end{equation}
i.e., $T=1$ corresponds to a periodic orbit [i.e., a periodic solution trajectory to the
dynamical system~\eqref{dynsys3E}], $\mathrm{L}$, with monotonically decreasing
$\theta$, and, hence, to a limit cycle for all solutions in the physical state space .

The structure on the past boundary is easily found since it consists
of two parts, one corresponding to an exponential potential and one to a constant potential.
It consists of fixed points and heteroclinic orbits (orbits that originate and end at distinct
fixed points) that join them. Since there are no heteroclinic cycles on the past boundary, it
follows that all interior physical orbits originate from the fixed points, which are given by
\begin{subequations}\label{fixedpoints}
\begin{alignat}{5}
\mathrm{M}_\pm\!: \quad \tilde{T} &= 0, &\quad T &= 0; &\quad \Sigma_\varphi &= \pm 1; &\quad X &= 0; &\quad \theta &= (2m \pm \sfrac12)\pi,\\
\mathrm{dS}\!: \quad \tilde{T} &= 1, &\quad T &= \sfrac12; &\quad \Sigma_\varphi &= 0; &\quad X &= 1; &\quad \theta &= 2m\pi,\\
\mathrm{PL}\!: \quad \tilde{T} &= 0, &\quad T &= 0; &\quad \Sigma_\varphi &= \bar{\lambda};
&\quad X &= - (1-\bar{\lambda}^2)^{\frac{1}{2n}}; &\quad
\theta &= \arccos X,
\end{alignat}
\end{subequations}
where $\mathrm{PL}$ only exists on the extended physical state space if $\bar{\lambda}<1$.
This fixed point corresponds to the self-similar solution for an exponential potential, which
yields a power-law solution, explaining the nomenclature. If $\bar{\lambda} < 1/\sqrt{3}$ this solution is accelerating
since $q = 3\bar{\lambda}^2 - 1$ for $\mathrm{PL}$. The nomenclature $\mathrm{M}_\pm$ corresponds to a
massless scalar field with $q=2$ (i.e., it corresponds to setting the potential to zero), where
$\mathrm{M}_\pm$ implies $\Delta\varphi = \pm\sqrt{6}N$, due to~\eqref{Sigdef}. Finally, $\mathrm{dS}$ stands for
de Sitter since $q=-1$ for this fixed point, although note that this de Sitter state corresponds to
$\varphi \rightarrow \infty$; i.e., it is an asymptotic state and not a physical de Sitter solution
with finite constant $\varphi$.

A local analysis of the fixed points shows that if $0<\bar{\lambda}<1$ then both $\mathrm{M}_+$ and
$\mathrm{M}_-$ are sources while $\mathrm{PL}$ is a saddle with a single solution entering the
physical state space. If $\bar{\lambda}\geq 1$ then $\mathrm{M}_-$ is a source, while $\mathrm{M}_+$
is a saddle from which no solutions enter the physical state space. Arguably, the most interesting
fixed point is $\mathrm{dS}$, which is a center saddle, with a one-dimensional center manifold
corresponding to the ``attractor solution'' or the ``inflationary trajectory''. To describe this
interior state space solution, which originates from $\mathrm{dS}$, we follow~\cite{alhugg15,alhetal15}
and perform a center manifold analysis. Since it is more convenient to use $\tilde{T}$ than $T$,
we use the system~\eqref{dynsys2E}, which results in the following center manifold expansion
(without loss of generality we choose $\theta = 0$ for $\mathrm{dS}$):
\begin{equation}\label{Emodasymptot}
\theta(\tilde{T}) = -\frac{\bar{\lambda}}{\sqrt{n}}(\tilde{T} - 1)
\left(1-\frac{\bar{\lambda}}{2\sqrt{n}}(\tilde{T} - 1) + \dots \right).
\end{equation}
The periodic orbit $\mathrm{L}$ corresponds to a blowup of the completely degenerate Minkowski
fixed point in the $(\dot{\varphi},\varphi)$ formulation. As discussed in~\cite{alhugg15,alhetal15},
to obtain explicit expressions for future asymptotics one can use available approximations for
late stage behavior when $V \sim \varphi^{2n}$, or one can use the averaging techniques developed
in~\cite{alhetal15}. The overall global solution structure for the $\mathrm{E}$-models
is depicted in Fig.~\ref{fig:Esolspace}. Note that $\mathrm{L}$ is the true future attractor,
and that it represents the future asymptotic behavior of \emph{all} physical solutions.

Finally, let us translate the above results to the original scalar field picture.
Let us first consider the inflationary solution coming from the $\mathrm{dS}$ fixed point.
In the vicinity of $\mathrm{dS}$, $\Sigma_\varphi = \frac{d\varphi}{dN}/\sqrt{6}$ is negative,
which means that toward the past, $\varphi$ is increasing. It is not difficult to
use the approximation~\eqref{Emodasymptot} to show that $\varphi$ to leading order can be 
written in the form $\varphi = A \ln(1-BN)$, where $A$ and $B$ are positive constants and 
where $N\rightarrow - \infty$ toward the past; i.e., the solution originates from 
$\varphi \rightarrow + \infty$ with a subsequently slowly decreasing $\varphi$, as expected.
Furthermore, note that initial data with an energy that is close to that
of an initial quasi-de Sitter state correspond to $T \approx \frac12$, where 
Fig.~\ref{fig:Esolspace} conveniently, at a glance, shows how solutions, and their properties, 
are distributed in terms of such initial data. From this figure it is obvious 
that there exists an open set with initial data with such initial energies that do not have a 
quasi-de Sitter phase in their evolution. Moreover, solutions that do have an 
intermediate de Sitter stage have solution trajectories that shadow the heteroclinic orbits 
$\mathrm{M}_\pm \rightarrow \mathrm{dS}$, which correspond to scalar field models with a 
constant potential $V_0$. 
 
From $\Sigma_\varphi = \frac{d\varphi}{dN}/\sqrt{6}$ it follows directly that $\mathrm{M}_-$
corresponds to a massless initial asymptotic state for which $\varphi \rightarrow +\infty$.
\emph{If} $0<\bar{\lambda} <1$, i.e., if $\alpha > (2n/3)^2$, then $\mathrm{M}_+$
corresponds to a massless initial asymptotic state with $\varphi \rightarrow -\infty$
from which an open set of solutions originates. Similarly, it follows that in this case 
the single solution that originates from $\mathrm{PL}$ corresponds to an initial power-law state
for which $\varphi \rightarrow -\infty$ toward the past; furthermore, if 
$\bar{\lambda} < 1/\sqrt{3}$ this is an initial power-law inflation state. 
Since $\mathrm{PL}$ is a saddle, it follows that there is an open set of solutions that
undergo an intermediate stage of power-law inflation in this case, but note that (i) this stage
is much shorter than the de Sitter stage since $\mathrm{PL}$ is a hyperbolic saddle in contrast 
to the center saddle $\mathrm{dS}$ and (ii) this inflationary stage takes place at a much larger energy scale
than that of the present quasi-de Sitter inflation. On the other hand, if $\bar{\lambda}\geq 1$,
i.e., if $\alpha \leq (2n/3)^2$, then all solutions originate from $\varphi \rightarrow +\infty$.
Finally, all solutions end at the Minkowski state associated with $\mathrm{L}$.

\section{$\mathrm{T}$-models}

\subsection{Dynamical systems formulations}

Using the dependent variables given in Eq.~\eqref{vardef}
and a new time variable $\tilde{\tau}$, defined by
\begin{equation}
\frac{d\tilde{\tau}}{dt} = H\tilde{T}^{-1},
\end{equation}
results in the following evolution equations for the state vector $(\tilde{T},\Sigma_\varphi,X)$,
\begin{subequations}\label{dynscalarT1}
\begin{align}
\frac{d\tilde{T}}{d\tilde{\tau}} &= \frac{3}{n}\Sigma_\varphi^2\tilde{T}^{2},\\
\frac{d\Sigma_\varphi}{d\tilde{\tau}} &= -3\left(\Sigma_\varphi X\tilde{T} + \frac12\bar{\lambda}(\tilde{T}^2 - X^2)\right)X^{2n-1},\\
\frac{dX}{d\tilde{\tau}} &= \frac{3}{n}\left(\Sigma_\varphi X\tilde{T} + \frac12\bar{\lambda}(\tilde{T}^2 - X^2)\right)\Sigma_\varphi,
\end{align}
and the constraint
\begin{equation}\label{Gauss3}
1 = \Sigma_\varphi^2 + X^{2n},
\end{equation}
\end{subequations}
where again
\begin{equation}
\bar{\lambda} = \frac{2n}{3\sqrt{\alpha}}.
\end{equation}

The constrained dynamical system~\eqref{dynscalarT1} admits a discrete symmetry
$(\Sigma_\varphi,X) \rightarrow -(\Sigma_\varphi,X)$, because the
potential is invariant under the transformation $\varphi \rightarrow -\varphi$.

The state space is bounded by the conditions that $\tilde{T}>0$ and that
\begin{equation}
\left(\frac{\tilde{T}}{\cosh\frac{\varphi}{\sqrt{6\alpha}}}\right)^2 = \tilde{T}^2 - X^2 >0.
\end{equation}
Since
\begin{equation}
\frac{d}{d\tilde{\tau}}(\tilde{T}^2 - X^2) =
\frac{6}{n}\Sigma_\varphi\left(\Sigma_\varphi\tilde{T} - \frac12\bar{\lambda}X\right)(\tilde{T}^2 - X^2),
\end{equation}
it follows that the physical state space is bounded toward the past by the invariant subset
$\tilde{T}^2 - X^2 =0$ for $\tilde{T} \geq 0$. Due to the discrete symmetry, this
invariant subset consists of two equivalent disconnected parts, one with $X>0$
and one with $X<0$, separated by the equivalent (massless state) fixed points
$\mathrm{M}_\pm$ at $X=0$, $\Sigma_\varphi=\pm 1$. The equations on the two branches of the
boundary subset, defined by $\tilde{T}^2 - X^2 =0$, are identical to those for a
constant potential. Thus, there are also two physically equivalent de Sitter
fixed points $\mathrm{dS}_\pm$ at $\Sigma_\varphi=0$, $X=\pm 1$. As for the previous
$\mathrm{E}$-models, we use the regularity of the equations to include the above boundary
subset in our analysis.

By solving the constraint using Eq.~\eqref{thetadef}, we obtain the following
dynamical system:
\begin{subequations}\label{dynscalar5}
\begin{align}
\frac{d\tilde{T}}{d\tilde{\tau}} &= \frac{3}{n}(1 - \cos^{2n}\theta)\tilde{T}^{2},\\
\frac{d\theta}{d\tilde{\tau}} &= -\frac{3}{2n}\left( G\tilde{T} \sin2\theta + \bar{\lambda}(\tilde{T}^2 - \cos^2{\theta})\right)G.
\end{align}
\end{subequations}
Changing $\tilde{T}$ to $T$ and the independent variable $\tilde{\tau}$ to $\check{\tau}$ according to
\begin{equation}
\frac{d\check{\tau}}{d\tilde{\tau}} = (1 + \tilde{T})^2 = (1-T)^{-2}
\end{equation}
results in
\begin{subequations}\label{dynscalar6}
\begin{align}
\frac{dT}{d\check{\tau}} &= \frac{3}{n}T^{2}(1-T)^2(1 - \cos^{2n}\theta),\\
\frac{d\theta}{d\check{\tau}} &= -\frac{3}{2n}\left(GT(1-T)\sin2\theta + \bar{\lambda}(T^2 - (1-T)^2 \cos^2{\theta})\right)G.
\end{align}
\end{subequations}

\begin{figure}[ht!]
\begin{center}
\includegraphics[width=0.45\textwidth]{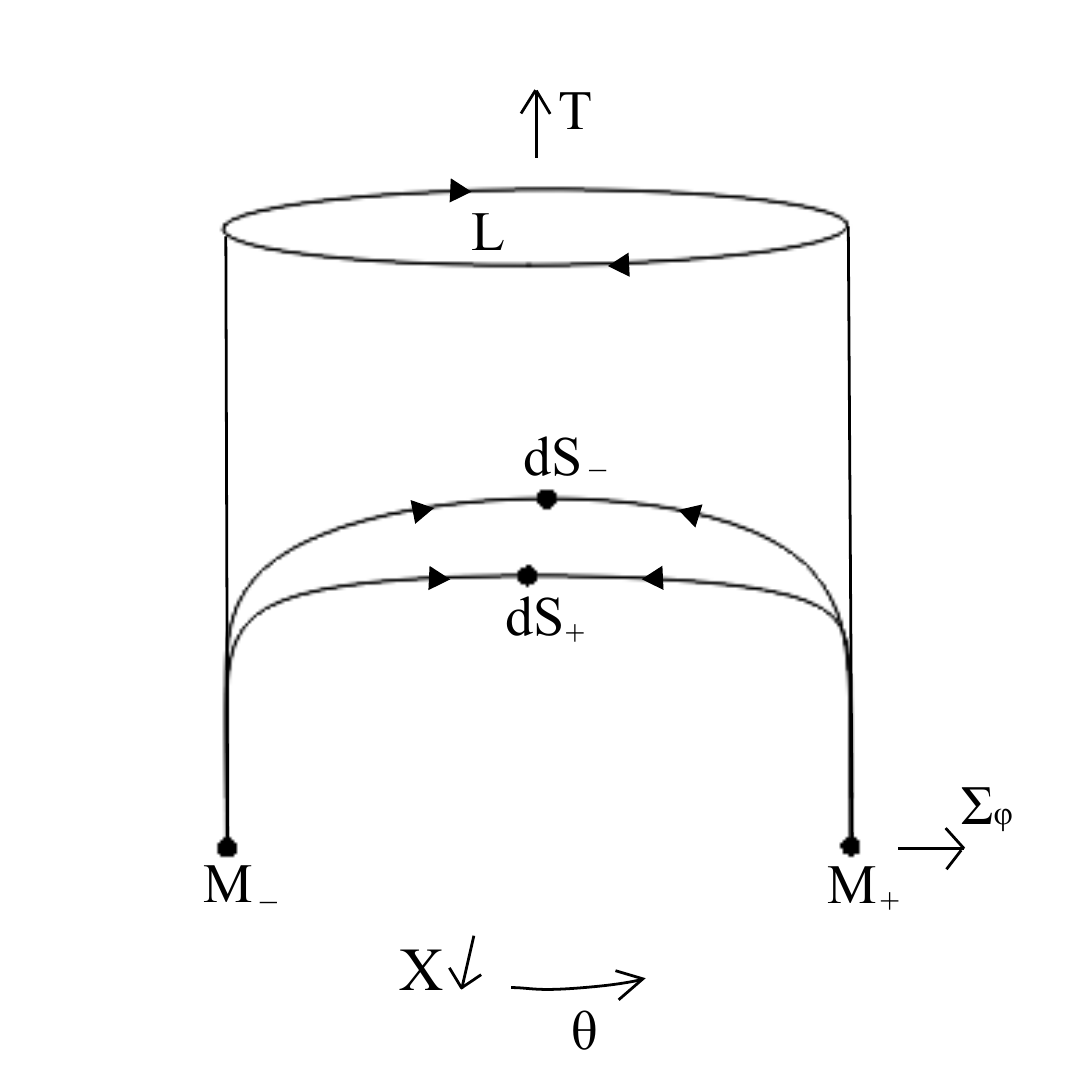}
\end{center}
\vspace{-0.5cm}
\caption{State space and boundary structures for $\mathrm{T}$-models with
$V = V_0\tanh^{2n}\frac{\varphi}{\sqrt{6\alpha}}$. 
}\label{fig:Tstatespace}
\end{figure}

Apart from including the past boundary, which in the present variables is given by
$T^2 - (1-T)^2 \cos^2{\theta} =0$, we also include the
future boundary $T=1$, which corresponds to $H=0$ and the final Minkowski
state. The resulting extended state space is therefore given by a finite cylinder
with the region $T^2 - (1-T)^2 \cos^2{\theta} < 0$ removed (see Fig.~\ref{fig:Tstatespace}).
%

\subsection{Dynamical systems analysis}

\begin{figure}[ht!]
\begin{center}
\subfigure[Solution space for the $\mathrm{T}$-models.]{\label{fig:}
\includegraphics[width=0.45\textwidth]{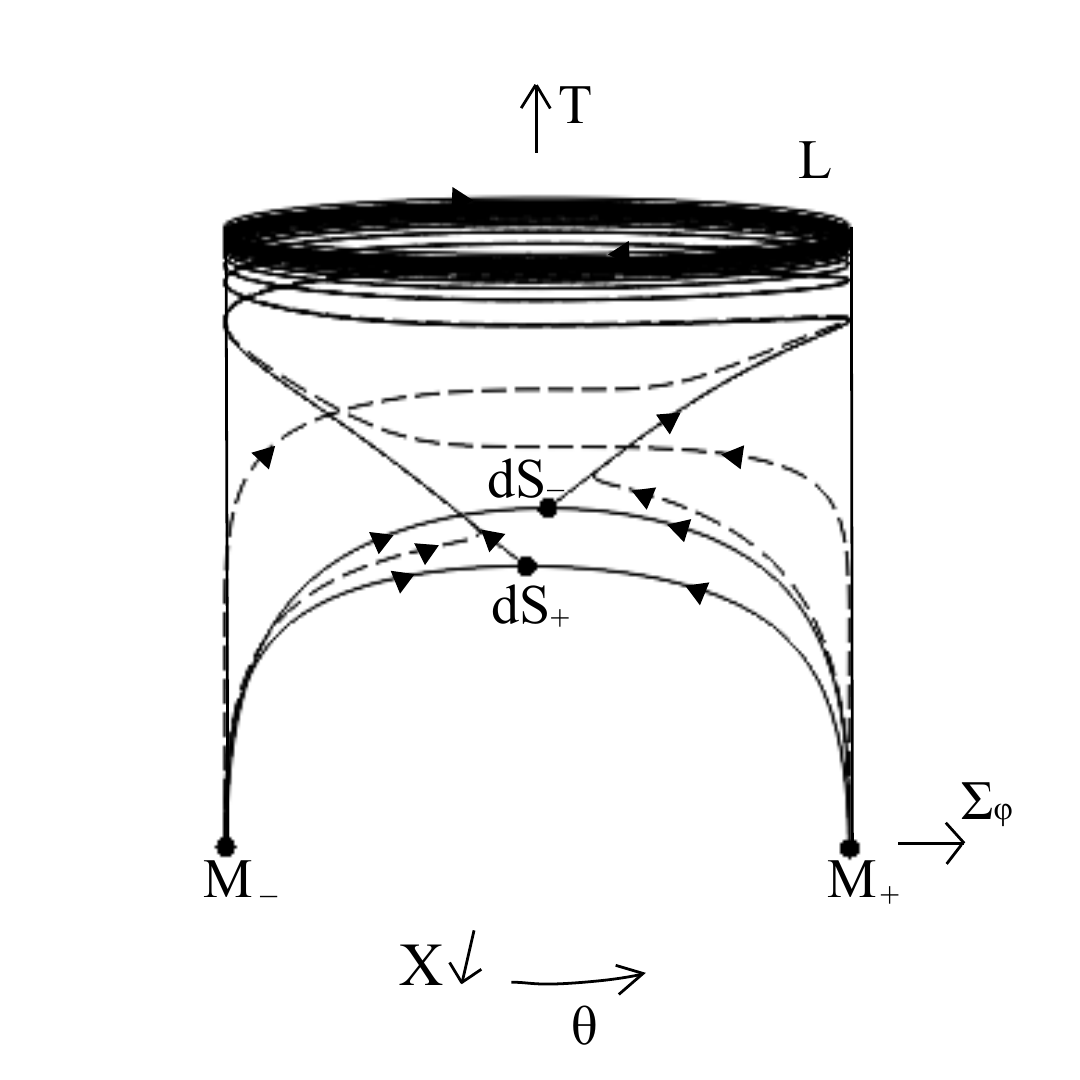}}
\subfigure[`Unwrapped' solution space for the $\mathrm{T}$-models.]{\label{fig:}
\includegraphics[width=0.45\textwidth]{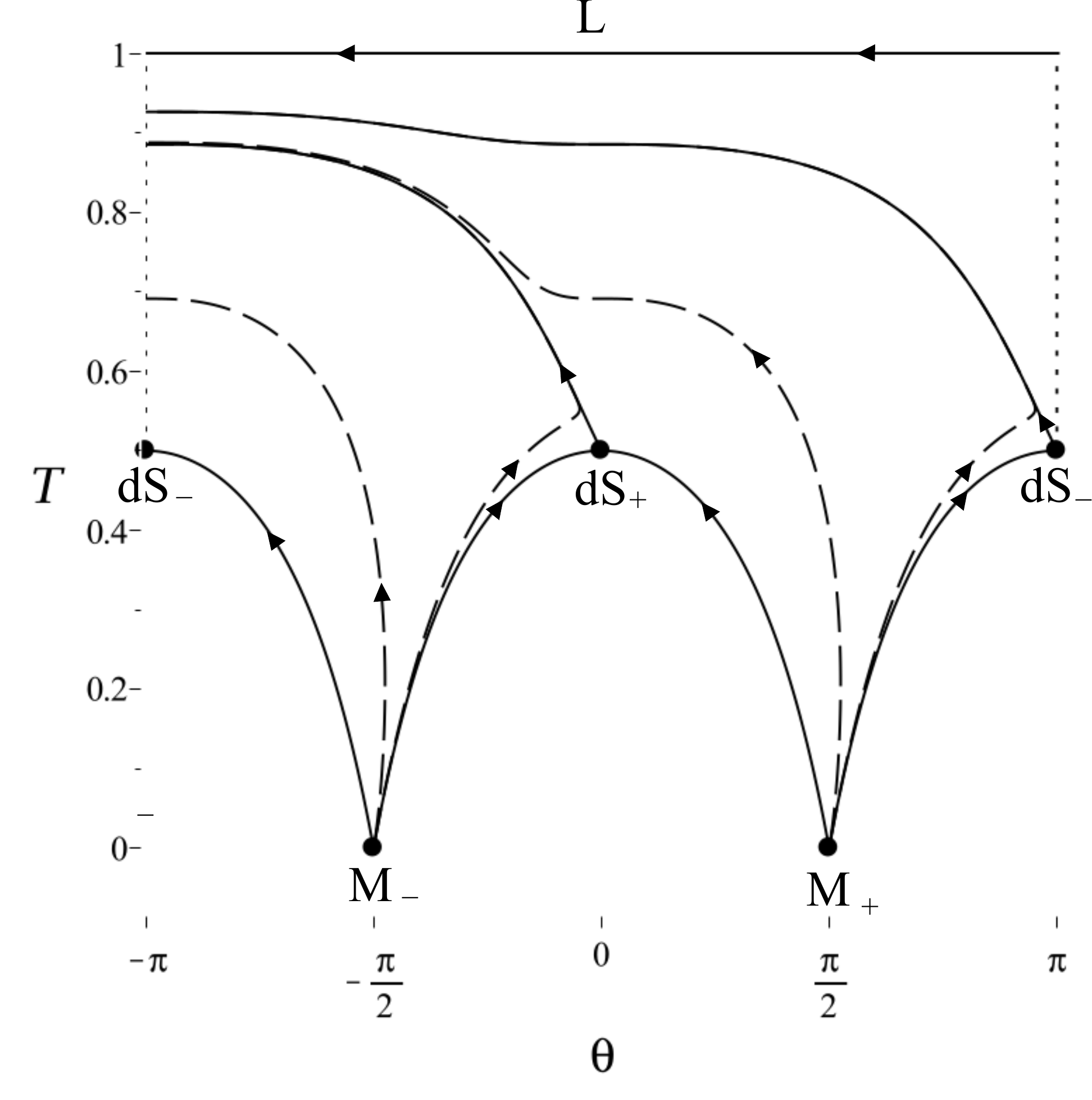}}
\end{center}
\vspace{-0.5cm}
\caption{
Representative solutions describing the solution space for the
$\mathrm{T}$-models with the potential
$V = V_0\tanh^{2n}\frac{\varphi}{\sqrt{6\alpha}}$
($n=1$, $\alpha =1$ in the figure).}
\label{fig:Tsolspace}
\end{figure}

As for the $\mathrm{E}$-models, since $H$ is monotonically decreasing, it follows
that $\tilde{T}$ and $T$ are monotonically increasing. Again, further insight
is gained by considering how $\tilde{T}$, and hence $T$, behave when
$\Sigma_\varphi=0\, \rightarrow\, \theta = m\pi$, where $q=-1$:
\begin{equation}\label{mon}
\left. \frac{d\tilde{T}}{d\tilde{\tau}}\right|_{q=-1} = 0,\qquad
\left. \frac{d^2\tilde{T}}{d\tilde{\tau}^2}\right|_{q=-1} = 0,\qquad
\left. \frac{d^3\tilde{T}}{d\tilde{\tau}^3}\right|_{q=-1} = \frac{27}{2n}\bar{\lambda}^2(\tilde{T} - 1)^2\tilde{T}^2.
\end{equation}
Since $\tilde{T} > 1$ when $q=-1$, it follows that
$\tilde{T}$, and hence $T$, is  monotonically increasing, although
the graphs of $\tilde{T}$ and $T$ go through inflection points when $q=-1$. Furthermore, since
\begin{equation}\label{mon}
\left. \frac{d\theta}{d\tilde{\tau}}\right|_{q=-1} = -\frac{3}{2n}\bar{\lambda}\left(\tilde{T}^2 - 1\right)G,
\end{equation}
it follows that $\theta$ is monotonically decreasing at $q=-1$, and thus that the solution
curves in the $T,\theta$ state space also for the $\mathrm{T}$ models become horizontal in
$T$ at $q=-1$ (see Fig.~\ref{fig:Tsolspace}).

From the monotonicity of $T$, and the discrete symmetry that makes the two fixed points
$\mathrm{M}_+$ and $\mathrm{M}_-$ physically equivalent, it follows that both these fixed points
are sources, corresponding to asymptotic massless self-similar states. The two
physically equivalent fixed points $\mathrm{dS}_+$ and $\mathrm{dS}_-$ are center saddles,
each yielding a single (physically equivalent) ``inflationary attractor solution'' entering
the physical state space, which, as for the $\mathrm{E}$-models, corresponds to a center manifold.
A center manifold expansion gives the following approximation for the inflationary attractor
solution (without loss of generality, we choose $\mathrm{dS}_+$ and $\theta=0$):
\begin{equation}
\theta(\tilde{T}) = -\frac{\bar{\lambda}}{\sqrt{n}}\left(\tilde{T} - 1\right)
\left(1-\frac{1}{2}\left(\frac{\bar{\lambda}^{2}}{n}+1\right)(\tilde{T}-1) + \dots \right).
\end{equation}

As in the $\mathrm{E}$-model case, the periodic orbit $\mathrm{L}$ corresponds to a blowup of the degenerate
Minkowski fixed point in the $(\dot{\varphi},\varphi)$ formulation, where $\mathrm{L}$ is the future
attractor.
The overall global solution structure is depicted in Fig.~\ref{fig:Tsolspace}. All
physical solutions originate from $\mathrm{M}_+$ and $\mathrm{M}_-$ (forming two
physically equivalent sets of solutions), apart from the two physically equivalent
inflationary attractor solutions that originate from $\mathrm{dS}_\pm$,
and all solutions end at the Minkowski state associated with the future attractor
and limit cycle $\mathrm{L}$.

In this case it is quite easy to translate the  results to the original scalar field picture.
A similar analysis to that for the $\mathrm{E}$-models shows that the open sets of physically 
equivalent solutions that originate from the massless states $\mathrm{M}_-$ and $\mathrm{M}_+$ 
correspond to initial states for which $\varphi \rightarrow + \infty$ and 
$\varphi \rightarrow -\infty$, respectively. The single solutions that come from the
de Sitter fixed points $\mathrm{dS}_+$ and $\mathrm{dS}_-$ correspond to the 
limits $\varphi \rightarrow +\infty$ and $\varphi \rightarrow -\infty$, respectively, 
from which they slowly evolve.  As in the $\mathrm{E}$-model case, all solutions end at the
Minkowski state associated with the limit cycle $\mathrm{L}$.

\section{Concluding Remarks}\label{sec:concl}

We begin this final section with some remarks on the relationship between
the center manifold and the slow-roll approximation for the inflationary
attractor solution. In the slow-roll approximation
$H\approx\sqrt{V(\varphi)/3}$ (i.e. $X=1$) is inserted into
$\dot{\varphi}=-2\frac{\partial H}{\partial \varphi}$, which gives
\begin{equation}
\dot{\varphi} \approx -\sqrt{\frac{V}{3}}\left(\frac{V_{\varphi}}{V}\right).
\end{equation}
In terms of $(\tilde{T},\Sigma_{\varphi},X)$, this yields the following
expressions for $\mathrm{E}$- and $\mathrm{T}$-models:
\begin{subequations}
\begin{align}
\Sigma_{\varphi} &\approx - \bar{\lambda}(\tilde{T} - X)X^{n-1}, \\
\Sigma_{\varphi} &\approx - \frac{\bar{\lambda}}{2}(\tilde{T} - \frac{X^2}{\tilde{T}})X^{n-1}.
\end{align}
\end{subequations}
In the vicinity of the asymptotic de Sitter state, where $\tilde{T} \approx 1$ and $\theta \approx 0$,
and therefore $\Sigma_{\varphi} \approx \sqrt{n}\theta$ [recall that $G(0) = \sqrt{n}$], $X \approx 1$,
these expressions yield
\begin{equation}
\theta(\tilde{T}) \approx -\frac{\bar{\lambda}}{\sqrt{n}}\left(\tilde{T} - 1\right),
\end{equation}
to lowest order. It follows that the slow-roll approximation leads to a curve in the $(T,\theta)$
state space that is tangential to the center manifold in the limit toward
the de Sitter state from which the center manifold, i.e. the inflationary attractor solution,
originates, as is also true for monomial potentials as discussed
in~\cite{alhugg15,alhetal15}.

The inflationary ``attractor'' solution, being a one-dimensional center manifold,
attracts nearby solutions exponentially rapidly, which then move along the center
manifold in a relatively slow power-law manner in the vicinity of the de Sitter
fixed point.\footnote{For a similar discussion in the context of quadratic theories
of gravity, see the paragraph after Eq. (20) in~\cite{barher06}.}
Thus, the center manifold structure explains both the attracting nature of the
inflationary ``attractor'' solution and the fact that nearby solutions obtain
a sufficient number of $e$-folds to be physically viable in an inflationary context.
Nevertheless, although this holds for an open set of solutions that shadow the past
boundary from fixed points that are sources to a de Sitter state on this boundary,
it should also be pointed out that there exists an open set of solutions that
behave differently, as seen in Figs.~\ref{fig:Esolspace} and~\ref{fig:Tsolspace}.
Ruling out these other solutions as physically irrelevant and explaining the
special role of the ``inflationary attractor solution'' beyond its center manifold
structure, thereby relies on paradigmatic assumptions relating the problem to broader
contexts. Examples of such contexts involve various proposed theoretical frameworks
as well as, e.g., scale considerations, illustrated in the discussion of 
e.g. Ref.~\cite{lin16b}, and various measures, motivated by, e.g., symplectic structures;
for a recent discussion on measures which might be applicable to the present models,
see~\cite{gruslo16}.

\subsection*{Acknowledgments}
A.A. is funded by the FCT Grant No. SFRH/BPD/85194/2012, and supported by Project No. PTDC/MAT-ANA/1275/2014,
and CAMGSD, Instituto Superior T{\'e}cnico by FCT/Portugal through UID/MAT/04459/2013.
C.U. would like to thank the CAMGSD, Instituto Superior T{\'e}cnico in Lisbon for kind hospitality.


\end{document}